\documentclass[prl,a4paper,twocolumn,superscriptaddress]{revtex4-1}

\usepackage[utf8x]{inputenc}
\usepackage{bm}
\usepackage{amssymb}
\usepackage{amsmath}
\usepackage{graphicx}
\usepackage{psfrag}
\usepackage{latexsym}
\usepackage{hyperref}

\newcommand{\del}{\partial}

\newcommand{\calo}{{\cal O}}

\newcommand{\dd}{{\delta}}

\newcommand{\eq}{\begin{equation}}
\newcommand{\eqx}{\end{equation}}
\newcommand{\eqn}{\begin{eqnarray}}
\newcommand{\eqnx}{\end{eqnarray}}

\newcommand{\ie}{{\it i.e.,}\ }

\newcommand{\cals}{{\cal S}}
\newcommand{\calh}{{\cal H} }
\newcommand{\cale}{{e} }
\newcommand{\calm}{{\cal M} }

\begin{document}

\title{Hot Big Bang from inflation without the inflaton reheating} 

\author{Alex Buchel}\email{abuchel@perimeterinstitute.ca}

\affiliation{\it Department of Physics and Astronomy,
University of Western Ontario, London, Ontario N6A 5B7, Canada}

\affiliation{\it Perimeter Institute for Theoretical Physics, Waterloo, Ontario N2L 2Y5, Canada}

\begin{abstract}

  According to the standard lore, a prolonged inflation
leaves a quantum field theory in a cold, low entropy state.
Thus, some mechanism is needed to reheat this post-inflationary
state, leaving a hot, thermal, radiation-dominated Universe.
Typically, reheating is achieved coupling the inflaton field
to the QFT degrees of freedom. We argue that the nonequilibrium
dynamics of a non-conformal QFT in (post-)inflationary background
space-time can produce hot quark-gluon plasma with the reheating temperature
of order the inflationary Hubble scale, without the inflaton coupling.

\end{abstract}

\maketitle

\emph{Introduction and summary.--} A common assumption is that the elementary particles populating
the Universe were created during the process of reheating
the Standard Model after inflation \cite{Linde:1987aa}.
The basic idea \cite{Kofman:1997yn}
is that the elementary particles are produced from
interactions with the oscillating inflaton.
The process of reheating completes when almost all the energy
of the inflaton is transferred to the thermal energy of
the Standard Model particles, and the inflaton settles in the minimum of its
effective potential. We have yet to observe the inflaton field,
therefore it is sensible to ask the question whether a post-inflationary
quantum field theory state can be intrinsically reheated
during the inflationary exit stage. In other words, can we decouple
the dynamics of the pre-Big Bang cosmological background space-time
from the reheating, thus alleviating the pressure on Particle
Physics to detect the inflaton in the collider experiments?

In this Letter we present precisely such scenario. The main
point is that a nonconformal QFT state, right before the exit from
the accelerated background state-time expansion, taken for simplicity here
to be a de Sitter space-time, is not a Bunch-Davies vacuum,
but rather is a dynamical fixed point (DFP)
\cite{Buchel:2017pto,Buchel:2021ihu}: it is characterized
by a constant  entropy production rate. Specifically,
if $a(t)$ is a scale factor of the QFT background $d$-dimensional
Friedmann-Lemaitre-Robertson-Walker (FLRW) Universe
\begin{equation}
ds_{d}^2=-dt^2+a(t)^2\ d{\bm x}^2\,,
\label{defflrm}
\end{equation}
the entropy production rate is given by
\begin{eqnarray}
\nabla\cdot \cals
&&= \frac{1}{a(t)^{d-1}}\ \frac{d}{dt}\left(a(t)^{d-1} s(t)\right)\nonumber\\
&&=
\frac{1}{a(t)^{d-1}}\ \frac{d}{dt} s_{comoving}(t)\ \ge\ 0\,,
\label{ds2}
\end{eqnarray}
where  the comoving $s_{comoving}$ and the physical $s$ entropy densities are
related as
\begin{equation}
s_{comoving}(t)\ =\ a(t)^{d-1}\ s(t)\,.
\label{defcomoving2}
\end{equation}
Specializing to de Sitter Universe, \ie $a(t)=e^{Ht}$,
in variety of strongly coupled
nonconformal QFTs with a holographic dual, the
entropy production rate was shown to be nonzero
\cite{Buchel:2017pto,Buchel:2017lhu,Buchel:2019pjb},
\begin{equation}
\nabla\cdot \cals\bigg|_{\rm de\ Sitter}=(d-1)\ H\ \ s_{ent}\ne 0\,,
\label{rate3}
\end{equation}
where the constant de Sitter physical entropy density $s(t)$
is called the {\it vacuum entanglement entropy} (VEE) density $s_{ent}$
\cite{Buchel:2017qwd}.

Consider now a simple model of exit from the accelerated expansion,
where the Hubble parameter $\calh$ evolves as 
\begin{equation}
\calh(t)\equiv \frac{d}{dt}\ln a(t)=\frac{H}{1+\exp(2\gamma t)}\,.
\label{exita}
\end{equation}
Here the constant energy scale $\gamma$ specifies the exit-rate from
the inflation, 
and the scale parameter $a(t)$ is normalized as $a(t=0)=1$.
For a rough estimate, in this model the exit from inflation
occurs during the time frame
$t\in\  \propto (-\gamma^{-1},\gamma^{-1})$, so that the scale factor
changes as
\begin{equation}
\ln a(t)\bigg|_{start}^{end}\equiv \ln\frac{a_e}{a_s}\ \sim\
\int_{-{\gamma^{-1}}}^{\gamma^{-1}} dt\
\frac{H}{1+\exp(2\gamma t)}=\frac{H}{\gamma}\,.
\label{dela}
\end{equation}
During the inflationary exit, the comoving entropy density
can only increase \eqref{ds2}, so
\begin{equation}
\underbrace{s_{comoving,s}}_{\sim a_s^{d-1} s_{ent}}\ \le\
\underbrace{s_{comoving,e}}_{\equiv a_e^{d-1} s_e} \,,
\label{dels}
\end{equation}
where we denoted by $s_e$ the 
physical entropy density of the post-inflationary QFT state. 
From \eqref{dels} we conclude that
\begin{equation}
s_e\ \sim\ s_{ent}\ \left(\frac{a_s}{a_e}\right)^{d-1}\ \sim s_{ent}\cdot
e^{-\frac{H}{\gamma}(d-1)} = s_{ent}\cdot \calo(1)\,.
\end{equation}
The post-inflationary state $ _e$ is nonequilibrium; its subsequent evolution
leads to its thermalization with the thermal entropy density
$s_{thermal} \ge s_e$. Since the equilibration of the post-inflationary
state occurs in a microcanonical ensemble, \ie at constant energy density,
from its  energy density $\cale=\cale_{e}$ at the inflationary exist,
and the equilibrium equation of state of the QFT plasma,
we can predict its thermalization - {\it the reheating} - temperature. 
In what follows we implement the outlined argument in a precise
holographic model. The use of the holographic correspondence
\cite{Maldacena:1997re} is a required computational tool
if we want to reliably discuss the nonequilibrium dynamics of strongly
coupled gauge theories.

\emph{The model.--} We consider a simple\footnote{Reheating in more realistic models of \cite{Buchel:2019qcq,Buchel:2019pjb,Buchel:2022sfx}
will be discussed elsewhere.} holographic toy model of a $2+1$-dimensional massive $QFT_3$ 
with the effective dual gravitational action\footnote{We set the radius $L$ of an asymptotic $AdS_4$ geometry 
to unity.}:
\begin{equation}
\begin{split}
S_4=\frac{1}{2\kappa^2}\int_{\calm_4} dx^4\sqrt{-\gamma}\left[R+6-\frac 12 \left(\nabla\phi\right)^2+\phi^2\right]\,.
\end{split}
\label{s4}
\end{equation}
The four dimensional gravitational constant $\kappa$ is related to the ultraviolet (UV) conformal fixed point
$CFT_3$  central charge $c$  as 
\begin{equation}
c=\frac{192}{\kappa^2}\,.
\label{c}
\end{equation}
$\phi$ is  a  gravitational  bulk scalar with 
\begin{equation}
L^2 m^2_\phi=-2 \;,
\label{mphi}
\end{equation}
which is dual to a dimension $\Delta_\phi=2$ operator $\calo_\phi$ of the boundary theory. 
$QFT_3$ is a relevant deformation of the UV  $CFT_3$ with 
\begin{equation}
\calh_{CFT}\ \to\ \calh_{QFT}= \calh_{CFT}+\Lambda\ \calo_\phi \;,
\label{defformation}
\end{equation}
with $\Lambda$ being the deformation mass scale. Thermodynamics of the boundary $QFT_3$ plasma
was discussed in \cite{Buchel:2009ge}, the thermalization of the theory in $R^{2,1}$ was studied in
\cite{Bosch:2017ccw}, and the de Sitter DFPs of the theory were analyzed in \cite{Buchel:2017lhu}.
de Sitter DFPs of the model discussed are stable \cite{Buchel:2017lhu,Buchel:2022hjz}.

We study $QFT_3$ dynamics in FLRW Universe with the Hubble parameter evolving as in
\eqref{exita}. 
A generic state of the boundary field theory with a gravitational dual \eqref{s4}, homogeneous and isotropic in the spatial
boundary coordinates $\boldsymbol{x}=\{x_1,x_2\}$, leads to a bulk gravitational metric ansatz
\begin{equation}
ds_4^2=2 dt\ (dr -A dt) +\Sigma^2\ d\boldsymbol{x}^2\,,
\label{EFmetric}
\end{equation}
with the warp factors $A,\Sigma$ as well as the bulk scalar $\phi$
depending only on $\{t,r\}$. From
the effective action \eqref{s4} we obtain the following equations of
motion:
\begin{equation}
\begin{split}
&0=d_+'\Sigma+d_+\Sigma\ \left(\ln\Sigma\right)'-\frac 32 \Sigma-\frac 14\Sigma \phi^2 \;,\\
&0=d_+'\phi+d_+\phi\ \left(\ln\Sigma\right)'+\frac{d_+\Sigma}{\Sigma}\ \phi'
+\phi\;,\\
&0=A''-2\frac{d_+\Sigma}{\Sigma^2}\ \Sigma'+\frac 12 d_+\phi\ \phi'\;,
\end{split}
\label{ev1}
\end{equation}
as well as the Hamiltonian constraint equation:
\begin{equation}
0=\Sigma''+\frac 14\Sigma  (\phi')^2\,,
\label{ham}
\end{equation}
and the momentum constraint equation:
\begin{widetext}
\begin{equation}
\begin{split}
&0=d_+^2\Sigma-2 A d_+'\Sigma-\frac{d_+\Sigma}{\Sigma^2}\
\left(A\Sigma^2\right)'
  +\frac 14\Sigma  \left((d_+\phi)^2 
 +2A\left(6+\phi^2 \right) \right)\,.
\end{split}
\label{mom}
\end{equation}
\end{widetext}
In \eqref{ev1}-\eqref{mom} 
we denoted $'= \frac{\del}{\del r}$, $\dot\ =\frac{\del}{\del t}$, 
and $d_+= \frac{\del}{\del t}+A \frac{\del }{\del r}$. 
The near-boundary $r\to\infty$ asymptotic behavior
of the metric
functions and the scalar encode the mass parameter $\Lambda$ and the boundary
metric scale factor $a(t)$:
\begin{eqnarray}
  &&\phi=\frac{\Lambda}{r}+\calo(r^{-2})\,, \Sigma=a\biggl({r}+\lambda+\calo(r^{-1})\biggr)\,, 
  \nonumber\\
&&A=\frac{r^2}{2}+\left(\lambda-\frac{\dot a }{a }\right)r+\calo(r^0)\,.
\label{bcdata}
\end{eqnarray}
$\lambda=\lambda(t)$ in \eqref{bcdata} is the residual radial coordinate diffeomorphism parameter \cite{Chesler:2013lia}.
An initial state of the boundary field theory is specified providing the scalar
profile $\phi(t_{init},r)$ and solving the
constraint \eqref{ham}, subject to the boundary
conditions \eqref{bcdata}. Equations \eqref{ev1} can then be used to evolve
the state.

The subleading terms in the boundary expansion of the
metric functions and the scalar encode the evolution of the  energy
density $\cale(t)$, the pressure $P(t)$ and the expectation values of the operator
$\calo_\phi(t)$ of the prescribed boundary QFT initial state.
\begin{widetext}
Specifically, extending the asymptotic expansion \eqref{bcdata} for $\{\phi, A\}$,
\begin{equation}
\begin{split}
&\phi=\frac{\Lambda}{r}+\frac{f_2(t)}{r^2}+\cdots\,,\ A=\frac{r^2}{2}+\left(\lambda-\frac{\dot a }{a }\right)r
+\frac{\lambda^2}{2}-\frac {\Lambda^2}{8}-\frac{\dot a}{a}\ \lambda-\dot\lambda
+\frac{1}{r}\left(\mu(t)-\frac \Lambda4 f_2(t)-\frac {\Lambda^2}{4}\lambda +\frac {\Lambda^2}{4}\ \frac{\dot a}{a} \right)
+\cdots\,,
\end{split}
\label{extension}
\end{equation}
the observables of interest can be computed following the
holographic renormalization of the model:
\begin{eqnarray}
&&2\kappa^2\ \cale(t)= -4\mu +\left(\dd_1\ \Lambda^3+2\dd_2\ \Lambda \frac{(\dot a)^2}{a^2}\right)\,,
\label{vev1}\\
&&2\kappa^2\ P(t)= -2\mu +\frac 12 \Lambda \left( f_2+\lambda \Lambda-\frac{\dot a}{a}\Lambda\right)+\left(-\dd_1\ \Lambda^3-2\dd_2\ \Lambda \frac{\ddot a}{a}\right)\,,
\label{vev2}\\
&&2\kappa^2\ \calo_\phi(t)=-f_2-\lambda\Lambda +\frac{\dot a}{a}\ \Lambda+\left(3 \dd_1\ \Lambda^2+\dd_2\ \left( 4 \frac{\ddot a}{a}+ 2\frac{(\dot a)^2}{a^2}\right) \right)\,,
\label{vev3}
\end{eqnarray}
where the terms in brackets, depending on arbitrary constants $\{\dd_1,\dd_2\}$, encode the renormalization scheme 
ambiguities.
\end{widetext}
Independent of the renormalization scheme, 
these expectation values 
satisfy the expected conformal Ward identity
\begin{equation}
\begin{split}
&-\cale+2P=-\Lambda \calo_\phi\,.
\end{split}
\label{ward}
\end{equation}
Furthermore, the conservation of the stress-energy tensor 
\begin{equation}\label{cons}
\frac{d\cale}{dt}+2\frac{\dot a}{a} (\cale+P)=0\,,
\end{equation}
is a consequence of the momentum constraint \eqref{mom}:   
\begin{equation}
\begin{split}
&0=\dot \mu+  \frac{\dot a }{a}\left(3\mu-\frac 14 \Lambda f_2\right)-\frac{\Lambda^2}{4}\
\frac{\dot a}{a}\ \left(\lambda-\frac{\dot a}{a}\right)\,. 
\end{split}
\label{ward2}
\end{equation}
From now on we choose a scheme with $\dd_i=0$.

One of the advantages of the holographic formulation of a QFT dynamics is the natural definition 
of its far-from-equilibrium entropy density. A gravitational geometry \eqref{EFmetric} 
has an apparent horizon located at $r=r_{AH}$, where \cite{Chesler:2013lia}
\begin{equation}
d_+\Sigma\bigg|_{r=r_{AH}}=0\,.
\label{defhorloc}
\end{equation} 
Following \cite{Booth:2005qc,Figueras:2009iu} we associate the non-equilibrium  entropy density $s$
of the boundary QFT  with the Bekenstein-Hawking entropy density of the apparent horizon  
\begin{equation}
a^2 s =\frac {2\pi}{\kappa^2}\ {\Sigma^2}\bigg|_{r=r_{AH}}\,.
\label{as}
\end{equation}
Using the holographic background equations of motion \eqref{ev1}-\eqref{mom} 
we find 
\begin{equation}
\frac{d(a^2 s)}{dt}=\frac{2\pi}{\kappa^2}\ (\Sigma^2)'\ \frac{
 (d_+\phi)^2}{\phi^2+6}\bigg|_{r=r_{AH}}\,.
\label{dasdt}
\end{equation}
Following \cite{Buchel:2017pto} it is easy to  prove that the entropy production rate as defined by \eqref{dasdt}
is non-negative, \ie 
\begin{equation}
\frac{d(a^2 s)}{dt}\ge 0\,,
\label{dasdt2}
\end{equation}
in holographic dynamics governed by \eqref{ev1}-\eqref{mom}.
We implement the holographic evolution as explained in \cite{Buchel:2021ihu}, adopting numerical
codes developed in \cite{Bosch:2017ccw,Buchel:2017map,Buchel:2017lhu,Buchel:2021ihu}.

Consider the dynamics of the system in the inflationary-exit time window, defined as
$\gamma t\in [-5,5]$. For $\gamma t \lesssim -5$, the model is in a de Sitter DFP\footnote{In practice we
initialize the system in an arbitrary spatially homogeneous and isotropic state at $Ht_{init}\ll -\frac{5}{\gamma}H$
and let it evolve
to an appropriate de Sitter DFP attractor, uniquely specified by the ratio $\frac{\Lambda}{H}$ \cite{Buchel:2017lhu}.}, and 
by $\gamma t \sim 5$ the Hubble parameter $\frac{\calh(t)}{H}\sim 4.5\cdot 10^{-5}$ is vanishingly small,
so that the subsequent dynamics is effectively in the Minkowski space-time with the constant scale factor
$a_e=a(t=5/\gamma)$ and the constant energy density $\cale_e=\cale(t=5/\gamma)$ (see \eqref{vev1} and \eqref{ward2}).
Thus, the evolution for $\gamma t> 5$ is just a thermalization of the post-inflationary state with
a fixed energy density $\cale_e$, which following \cite{Bosch:2017ccw} would thermalize as $\gamma t\gg 1$ to
an equilibrium state with $\cale_{thermo}=\cale_e$. The thermal features of the final equilibrium state, including
its temperature --- {\it the reheating temperature we are after} --- can be read off from the
equation of state of the equilibrium quark-gluon plasma of the model determined in \cite{Buchel:2009ge}. 

\begin{figure}
\includegraphics[width=0.47\textwidth]{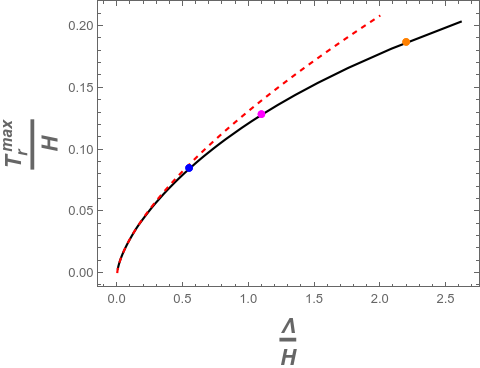}
\caption{Maximal reheating temperature $T_r^{max}$ in the nonconformal
holographic model \eqref{s4} with the mass scale $\Lambda$ \eqref{defformation},
evolving in FLRW cosmology with the Hubble parameter specified by \eqref{exita}
in the inflationary rapid-exit limit $\frac{H}{\gamma}\to 0$.
The red dashed curve is the near-conformal approximation to the maximal reheating temperature,
see \eqref{tpert}. The dots represent select value of $\frac{\Lambda}{H}$ for which we present the
reheating temperature for finite $\frac{\gamma}{H}$ (see~Fig.~\ref{figure2}).
}
\label{figure1}
\end{figure}

\begin{figure}
\includegraphics[width=0.47\textwidth]{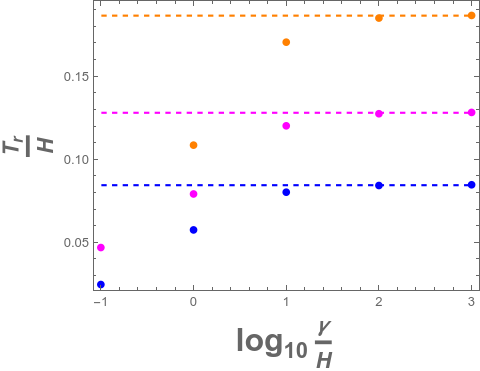}
\caption{The reheating temperature $T_r$  in the nonconformal
  holographic model \eqref{s4} as a function of $\log_{10}\frac{\gamma}{H}$ for select values of
  $\frac{\Lambda}{H}$. Horizontal dashed lines represent the maximal reheating temperature, achieved
  in the inflationary rapid-exit limit $\frac{H}{\gamma}\to 0$  (see~Fig.~\ref{figure1}).
}
\label{figure2}
\end{figure}

We find\footnote{Additional details regarding the simulations can be found
in the supplemental material.} that the maximal reheating temperature occurs when the exit from the de Sitter (inflationary) phase is
very rapid $\frac {\gamma}{H}\gg 1$. In this regime effectively all of the energy density $\cale_{DFP}$
of the pre-exit DFP state
is converted into the energy density $\cale_e$.
In the limit $\frac {\gamma}{H}\to \infty$ there is effectively no expansion of the Universe in the exit time-window (see \eqref{dela}), so there is no
'dilution' of the energy density. In this parametric regime, there is a clear separation of two processes:
the exit from the accelerated expansion, and the following equilibration of the post-exit state. The time
scale of the former is set by $\gamma^{-1}$, while the time scale of the latter is determined
by the thermalization (reheating) temperature $T_r$ \cite{Buchel:2015saa,Buchel:2015ofa,Attems:2016ugt,Janik:2016btb}.
For $\gamma \lesssim H$ the two processes become intertwined, the scale factor of the Universe can noticeably
increase, resulting in the dilution of the energy density leading to $\cale_e<\cale_{DFP}$, and ultimately decreasing the
reheating temperature. Results of the numerical analysis are presented in Figs.~\ref{figure1},\ref{figure2}.
The red dashed curve in the left panel represents the near-conformal, \ie $\Lambda\ll H$, approximation
to the maximal reheating temperature\footnote{It can be analytically computed
using the near-conformal description of the de Sitter DFPs of the model \cite{Buchel:2017lhu}
and its equilibrium thermodynamics. },
\begin{equation}
\frac{T_r^{max}}{H}\approx \frac{3^{2/3}}{2^{7/3}\pi} \left(\frac{\Lambda}{H}\right)^{2/3}\,.
\label{tpert}
\end{equation}
Notice that the reheating temperature vanishes in the conformal limit --- (unjustified) expectations that
non-conformal theories have trivial de Sitter vacua, rather than de Sitter DFPs, was the prime reason
for the inflaton reheating models \cite{Kofman:1997yn}.

\vspace{5 pt}

\emph{Conclusions.--} We argued that the reheating of a post-inflationary state of a nonconformal quantum
field theory can be achieved entirely due to the  nonequilibrium dynamics in the inflationary exit.
The reheating temperature is the larger the more rapid the exit from inflation occurs, and the larger
the scale invariance of the QFT is broken compare to the inflationary Hubble scale $H$. In the model
discussed we observed the reheating temperatures of order $T_r\sim \frac {1}{10} H$ for the mass parameter
of the theory $\Lambda \sim H$.

The main open problem is understanding de Sitter DFPs from purely QFT perspective.
Whenever the exit from the accelerated expansion is very rapid, the pre-thermalized state of the theory is very close to that
of the corresponding DFP; this universality suggests that in this case there might be observable phenomenological imprints of the
dynamical fixed point of the theory on the Hot Big Bang cosmology.

The reheating mechanism described in this Letter is that of the gravitational
reheating\footnote{See \cite{Akrami:2017cir} for a recent review and additional references.}.
In this sense it is similar to preheating in "non-oscillatory models'' (NO) \cite{Peebles:1998qn,Felder:1999pv}.
The difference is related to the question,
{\it what is the energy being released} upon the exit from inflation? This
is the crucial issue, since
once the energy is released, the corresponding QFT state will simply
equilibrate to the thermal one at this particular energy density.
The energy released at the end of inflation in NO models is the energy
of the produced particles during the inflationary exit. On the contrary, here, it
is the energy density of the de Sitter state of the
strongly coupled non-conformal theory that is being released --- very
little energy density is being produced during the (rapid)
inflationary exit.
In fact, the more rapid the exit, the less additional energy is
being produced (see the supplemental material).

\vspace{5 pt}

\begin{acknowledgements}
  We would like to thank A. Linde for valuable correspondence
  regarding preheating in NO models.
  Research at Perimeter Institute is supported in part by the Government
of Canada through the Department of Innovation, Science and Economic
Development Canada and by the Province of Ontario through the Ministry
of Colleges and Universities. This work is further supported by a
Discovery Grant from the Natural Sciences and Engineering Research
Council of Canada.

\end{acknowledgements}

\vspace{10 pt}

\emph{Supplemental material.--} We use numerical code developed in
\cite{Buchel:2017lhu,Buchel:2021ihu} to evolve the holographic model
\eqref{s4} in FLRW Universe \eqref{exita} --- the modifications are minimal
as the evolution equations \eqref{ev1} are sensitive only
to $\frac{\calh(t)}{H}$ (and not its derivatives),
which is set to unity in \cite{Buchel:2017lhu,Buchel:2021ihu} for the
study of the de Sitter DFPs of the model. This change neither
affects the convergence nor the stability of the code for
inflationary exit simulations over the wide range of 
$\frac{\gamma}{H}=10^{-1}\cdots 10^3$. In what follows we present
main simulation results to collaborate the statements in the Letter.
We focus on simulations with $\frac{\Lambda}{H}=0.55$.

\begin{figure}
\includegraphics[width=0.47\textwidth]{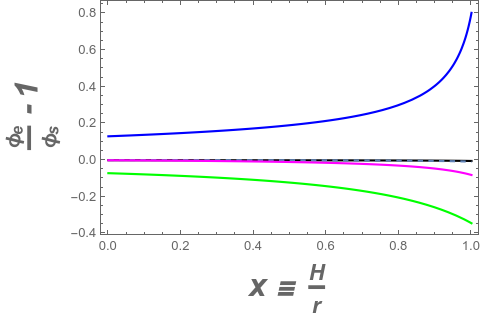}
\caption{Comparison of the bulk scalar profiles
  $\phi_s=\phi(\gamma t=-5)$ and $\phi_e=\phi(\gamma t=5)$
  over the full computational domain $x\in [0,1]$ for
  different inflationary exit rates $\frac{\gamma}{H}$
  (see \eqref{color}).
}
\label{figure3}
\end{figure}

\begin{figure}
\includegraphics[width=0.47\textwidth]{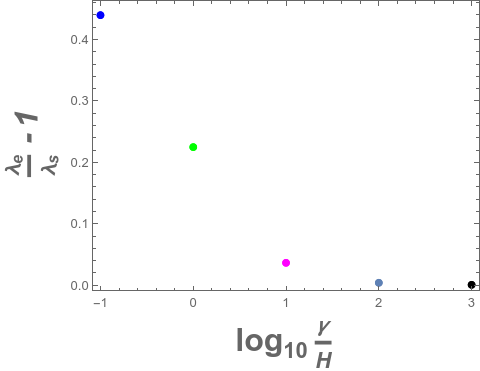}
\caption{Comparison of the diffeomorphism parameter
  $\lambda$ at the start of the exit $\gamma t =-5$, and
  at the end of the exit $\gamma t =5$, for
  different inflationary exit rates $\frac{\gamma}{H}$
  (see \eqref{color}).
}
\label{figure4}
\end{figure}

In Figs.~\ref{figure3},\ref{figure4} we present the relative change 
of the gravitational bulk scalar $\phi$ and
the diffeomorphism parameter $\lambda$ (see \eqref{extension})
during the inflationary exit for select values of
$\frac{\gamma}{H}$. Here,
the profile of the scalar at the inflationary exit {\it start},
$\phi_s\equiv \phi(t=-\frac{5}{\gamma},r)$, is identical to that of
corresponding de Sitter dynamical fixed point;
while $\phi_e\equiv \phi(t=+\frac{5}{\gamma},r)$ indicates the
scalar profile at the {\it end} of the inflationary exit.
Similar notations are used for the parameter $\lambda$.
We use consistent color coding throughout the supplemental material,
so that
\begin{equation}
 \log_{10} \frac{\gamma}{H}=\biggl\{\ \underbrace{-1}_{\rm blue}\,,\,
 \underbrace{0}_{\rm green}\,,\, \underbrace{1}_{\rm magenta}\,,\,
 \underbrace{2}_{\rm grey\ dashed}\,,\, \underbrace{3}_{\rm black}
  \ \biggr\}\,.
  \label{color}
  \end{equation}
The main message is that the more rapid the exit from inflation is,
\ie the larger is the ratio $\frac{\gamma}{H}$, the more
the exit state $ _e$ resembles that of the pre-exit de Sitter
dynamical fixed point\footnote{This is a different universality
from the abrupt holographic quenches of the relevant couplings
of the boundary QFT studied in \cite{Buchel:2013gba}.}. 

\begin{figure}
\includegraphics[width=0.47\textwidth]{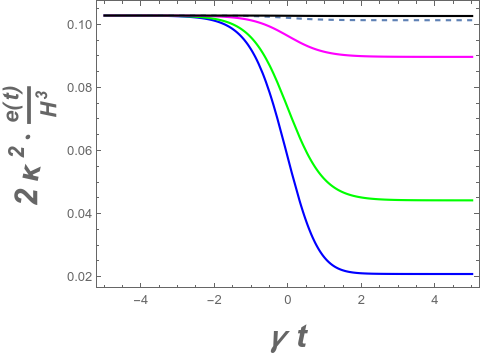}
\caption{The energy density profile $e(t)$ (see \eqref{vev1})
  for inflationary exits with different rates $\frac{\gamma}{H}$
  (see \eqref{color}).
}
\label{figure5}
\end{figure}

Thus, it is not a surprise that there is less energy density
change in the inflationary exit, the more abrupt
this exit is, see Fig.~\ref{figure5}. At the start of the exit,
\ie for $\gamma t \sim -5$, the energy density is that of the
de Sitter DFP of the holographic model \eqref{s4} with
$\frac{\Lambda}{H}=0.55$. Notice that the physical energy density
decreases for more gradual exit rates (the blue curve) --- this is
because for smaller values of $\frac{\gamma}{H}$ the Universe can substantially
expand in the exit window $\gamma t\in [-5,5]$. 

\begin{figure}
\includegraphics[width=0.47\textwidth]{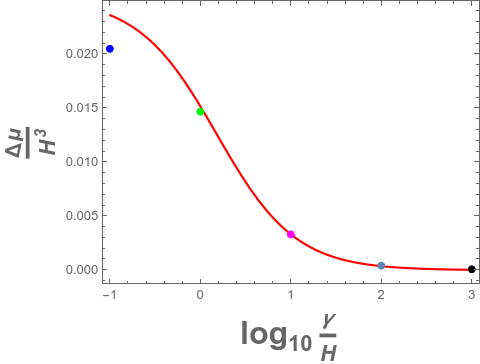}
\caption{The energy density change $2\kappa^2\Delta\cale \equiv -4\Delta\mu $
  (see \eqref{vev1})
  for inflationary exits with different rates $\frac{\gamma}{H}$
  (see \eqref{color}). The solid red curve is the
  'abrupt exit' approximation \eqref{delmu}.  
}
\label{figure6}
\end{figure}

As an important check on our numerics,
we can analytically predict the energy density change in the
inflationary exit window in the abrupt exit limit.
Indeed, using the fact that in the limit $\gamma\gg H$
both $f_2$ and $\lambda$ are nearly constant,
\begin{equation}
  f_2(t)\approx f_2^{DFP}\,,\qquad \lambda(t)\approx \lambda^{DFP}\,,
  \label{rexit}
  \end{equation}
we can integrate \eqref{ward2}:
\begin{widetext}
\begin{equation}
  \mu(t)=\underbrace{\frac{\Lambda}{12}(f_2^{DFP}+H\lambda^{DFP}-H\Lambda)
  }_{=\mu^{DFP}}+
  \frac{\Lambda^2 H}{4}\cdot \frac{1}{2\frac{\gamma}{H}+3}\cdot
  \frac{1}{1+\exp(-2\gamma t)}\,.
  \label{muan}
  \end{equation}
\end{widetext}
Thus, as $\frac{H}{\gamma}\to 0$ and
dropping $\exp(-2\gamma t_e)=\exp(-10)\sim 0$, we find
\begin{equation}
  \frac{\Delta\mu}{H^3}\equiv \frac{\mu_e-\mu^{DFP}}{H^3}\approx
  \frac{\Lambda^2}{4H^2}\cdot \frac{1}{2\frac{\gamma}{H}+3}
  \label{delmu}
  \end{equation}
Results for $\frac{\Delta\mu}{H^3}$ for inflationary exits
with rates \eqref{color}, along with the analytic approximation
(the solid red curve) \eqref{delmu} are presented in
Fig.~\ref{figure6}.

\begin{figure}
\includegraphics[width=0.47\textwidth]{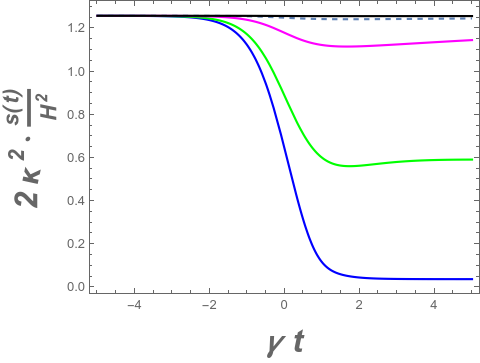}
\caption{The entropy density profile $s(t)$ (see \eqref{as})
  for inflationary exits with different rates $\frac{\gamma}{H}$
  (see \eqref{color}).
}
\label{figure7}
\end{figure}

\begin{figure}
\includegraphics[width=0.47\textwidth]{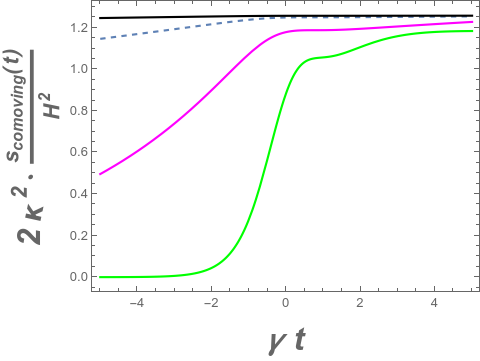}
\caption{The comoving entropy density profile $s_{comoving}(t)$
  (see \eqref{defcomoving2})
  for inflationary exits with different rates $\frac{\gamma}{H}$
  (see \eqref{color}).
}
\label{figure8}
\end{figure}

In Figs.~\ref{figure7},\ref{figure8} we present the
evolution of the physical $s(t)$ and the comoving $s_{comoving}(t)$
entropy density of our holographic model in the inflationary
exits with rates\footnote{We do not provide the plot for
the comoving entropy density for the evolution with
$\frac{\gamma}{H}=10^{-1}$. Given the vast expansion of the Universe
in this case, $\ln\frac{a_e}{a_s}=50$,
it would completely overwhelm the other results.
} given by \eqref{color}. Note that at the start of the inflationary
exits, the physical entropy density $s_s\equiv
s(t= t_s=-\frac{5}{\gamma})$ is the same for all the exit rates. It
is nothing but the vacuum entanglement entropy density
$s_{ent}$ (see \eqref{rate3})
of the corresponding de Sitter DFP of the model. On the contrary, the
comoving entropy densities are very different: $s_{comoving,s}=a(t_s)^2 \cdot s_{ent}$,
where the scale factor $a(t_s)$ is
\begin{equation}
  a(t_s)=\left(\frac{1}{1+\exp(-2\gamma t)}\right)^{\frac{H}{2\gamma}}\bigg|_{t=t_s=-\frac{5}{\gamma}}
  \label{af}
  \end{equation}
The physical entropy density initially decreases
(due to the continual expansion of the Universe in the inflationary exit),
but close to the end of the 
inflationary exit, \ie for $\gamma t\sim 5$,
it starts to increase, representing the thermalization
of the inflationary post-exit state of the model in Minkowski space-time.
Following \eqref{dasdt2}, the comoving entropy density always increases 
throughout the full dynamics. 

\begin{figure}
\includegraphics[width=0.47\textwidth]{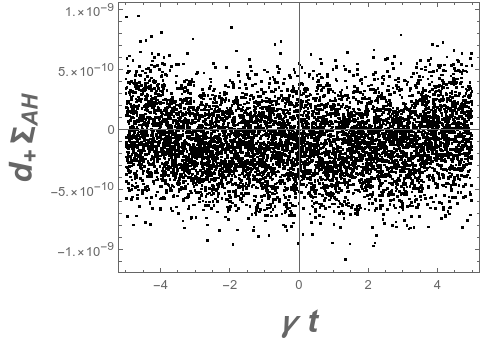}
\caption{We monitor the constraint \eqref{defhorloc}
  to validate the code. 
}
\label{figure9}
\end{figure}

The final comment is related to one of the many checks we performed on
our numerical code. During the evolution, the computational domain stays
fixed, $x\equiv \frac{H}{r}\in [0,1]$, with $x\to 0$ corresponding to
the asymptotic $AdS_4$ boundary, and $x\to 1$ representing the apparent
horizon of the dual gravitational bulk. An important test is monitoring 
the constraint \eqref{defhorloc} throughout the full evolution.
In Fig.~\ref{figure9} we present $d_+\Sigma$, evaluated
at the apparent horizon, during the inflationary exit with
$\frac{\gamma}{H}=10^3$.

\bibliography{dsexit_biblio}{}

\begin{thebibliography}{33}%
\makeatletter
\providecommand \@ifxundefined [1]{%
 \@ifx{#1\undefined}
}%
\providecommand \@ifnum [1]{%
 \ifnum #1\expandafter \@firstoftwo
 \else \expandafter \@secondoftwo
 \fi
}%
\providecommand \@ifx [1]{%
 \ifx #1\expandafter \@firstoftwo
 \else \expandafter \@secondoftwo
 \fi
}%
\providecommand \natexlab [1]{#1}%
\providecommand \enquote  [1]{``#1''}%
\providecommand \bibnamefont  [1]{#1}%
\providecommand \bibfnamefont [1]{#1}%
\providecommand \citenamefont [1]{#1}%
\providecommand \href@noop [0]{\@secondoftwo}%
\providecommand \href [0]{\begingroup \@sanitize@url \@href}%
\providecommand \@href[1]{\@@startlink{#1}\@@href}%
\providecommand \@@href[1]{\endgroup#1\@@endlink}%
\providecommand \@sanitize@url [0]{\catcode `\\12\catcode `\$12\catcode
  `\&12\catcode `\#12\catcode `\^12\catcode `\_12\catcode `\%12\relax}%
\providecommand \@@startlink[1]{}%
\providecommand \@@endlink[0]{}%
\providecommand \url  [0]{\begingroup\@sanitize@url \@url }%
\providecommand \@url [1]{\endgroup\@href {#1}{\urlprefix }}%
\providecommand \urlprefix  [0]{URL }%
\providecommand \Eprint [0]{\href }%
\providecommand \doibase [0]{http://dx.doi.org/}%
\providecommand \selectlanguage [0]{\@gobble}%
\providecommand \bibinfo  [0]{\@secondoftwo}%
\providecommand \bibfield  [0]{\@secondoftwo}%
\providecommand \translation [1]{[#1]}%
\providecommand \BibitemOpen [0]{}%
\providecommand \bibitemStop [0]{}%
\providecommand \bibitemNoStop [0]{.\EOS\space}%
\providecommand \EOS [0]{\spacefactor3000\relax}%
\providecommand \BibitemShut  [1]{\csname bibitem#1\endcsname}%
\let\auto@bib@innerbib\@empty
\bibitem [{\citenamefont {Linde}(1987)}]{Linde:1987aa}%
  \BibitemOpen
  \bibfield  {author} {\bibinfo {author} {\bibfnamefont {A.~D.}\ \bibnamefont
  {Linde}},\ }\href {\doibase 10.1063/1.881088} {\bibfield  {journal} {\bibinfo
   {journal} {Phys. Today}\ }\textbf {\bibinfo {volume} {40}},\ \bibinfo
  {pages} {61} (\bibinfo {year} {1987})}\BibitemShut {NoStop}%
\bibitem [{\citenamefont {Kofman}\ \emph {et~al.}(1997)\citenamefont {Kofman},
  \citenamefont {Linde},\ and\ \citenamefont {Starobinsky}}]{Kofman:1997yn}%
  \BibitemOpen
  \bibfield  {author} {\bibinfo {author} {\bibfnamefont {L.}~\bibnamefont
  {Kofman}}, \bibinfo {author} {\bibfnamefont {A.~D.}\ \bibnamefont {Linde}}, \
  and\ \bibinfo {author} {\bibfnamefont {A.~A.}\ \bibnamefont {Starobinsky}},\
  }\href {\doibase 10.1103/PhysRevD.56.3258} {\bibfield  {journal} {\bibinfo
  {journal} {Phys. Rev. D}\ }\textbf {\bibinfo {volume} {56}},\ \bibinfo
  {pages} {3258} (\bibinfo {year} {1997})},\ \Eprint
  {http://arxiv.org/abs/hep-ph/9704452} {arXiv:hep-ph/9704452} \BibitemShut
  {NoStop}%
\bibitem [{\citenamefont {Buchel}\ and\ \citenamefont
  {Karapetyan}(2017)}]{Buchel:2017pto}%
  \BibitemOpen
  \bibfield  {author} {\bibinfo {author} {\bibfnamefont {A.}~\bibnamefont
  {Buchel}}\ and\ \bibinfo {author} {\bibfnamefont {A.}~\bibnamefont
  {Karapetyan}},\ }\href {\doibase 10.1007/JHEP03(2017)114} {\bibfield
  {journal} {\bibinfo  {journal} {JHEP}\ }\textbf {\bibinfo {volume} {03}},\
  \bibinfo {pages} {114} (\bibinfo {year} {2017})},\ \Eprint
  {http://arxiv.org/abs/1702.01320} {arXiv:1702.01320 [hep-th]} \BibitemShut
  {NoStop}%
\bibitem [{\citenamefont {Buchel}(2022{\natexlab{a}})}]{Buchel:2021ihu}%
  \BibitemOpen
  \bibfield  {author} {\bibinfo {author} {\bibfnamefont {A.}~\bibnamefont
  {Buchel}},\ }\href {\doibase 10.1007/JHEP02(2022)128} {\bibfield  {journal}
  {\bibinfo  {journal} {JHEP}\ }\textbf {\bibinfo {volume} {02}},\ \bibinfo
  {pages} {128} (\bibinfo {year} {2022}{\natexlab{a}})},\ \Eprint
  {http://arxiv.org/abs/2111.04122} {arXiv:2111.04122 [hep-th]} \BibitemShut
  {NoStop}%
\bibitem [{\citenamefont {Buchel}(2018)}]{Buchel:2017lhu}%
  \BibitemOpen
  \bibfield  {author} {\bibinfo {author} {\bibfnamefont {A.}~\bibnamefont
  {Buchel}},\ }\href {\doibase 10.1016/j.nuclphysb.2018.01.021} {\bibfield
  {journal} {\bibinfo  {journal} {Nucl. Phys.}\ }\textbf {\bibinfo {volume}
  {B928}},\ \bibinfo {pages} {307} (\bibinfo {year} {2018})},\ \Eprint
  {http://arxiv.org/abs/1707.01030} {arXiv:1707.01030 [hep-th]} \BibitemShut
  {NoStop}%
\bibitem [{\citenamefont {Buchel}(2020)}]{Buchel:2019pjb}%
  \BibitemOpen
  \bibfield  {author} {\bibinfo {author} {\bibfnamefont {A.}~\bibnamefont
  {Buchel}},\ }\href {\doibase 10.1007/JHEP05(2020)035} {\bibfield  {journal}
  {\bibinfo  {journal} {JHEP}\ }\textbf {\bibinfo {volume} {05}},\ \bibinfo
  {pages} {035} (\bibinfo {year} {2020})},\ \Eprint
  {http://arxiv.org/abs/1912.03566} {arXiv:1912.03566 [hep-th]} \BibitemShut
  {NoStop}%
\bibitem [{\citenamefont {Buchel}(2017{\natexlab{a}})}]{Buchel:2017qwd}%
  \BibitemOpen
  \bibfield  {author} {\bibinfo {author} {\bibfnamefont {A.}~\bibnamefont
  {Buchel}},\ }\href@noop {} {\  (\bibinfo {year} {2017}{\natexlab{a}})},\
  \Eprint {http://arxiv.org/abs/1702.08590} {arXiv:1702.08590 [hep-th]}
  \BibitemShut {NoStop}%
\bibitem [{\citenamefont {Maldacena}(1999)}]{Maldacena:1997re}%
  \BibitemOpen
  \bibfield  {author} {\bibinfo {author} {\bibfnamefont {J.~M.}\ \bibnamefont
  {Maldacena}},\ }\href {\doibase 10.1023/A:1026654312961,
  10.4310/ATMP.1998.v2.n2.a1} {\bibfield  {journal} {\bibinfo  {journal} {Int.
  J. Theor. Phys.}\ }\textbf {\bibinfo {volume} {38}},\ \bibinfo {pages} {1113}
  (\bibinfo {year} {1999})},\ \bibinfo {note} {[Adv. Theor. Math.
  Phys.2,231(1998)]},\ \Eprint {http://arxiv.org/abs/hep-th/9711200}
  {arXiv:hep-th/9711200 [hep-th]} \BibitemShut {NoStop}%
\bibitem [{Note1()}]{Note1}%
  \BibitemOpen
  \bibinfo {note} {Reheating in more realistic models of \cite
  {Buchel:2019qcq,Buchel:2019pjb,Buchel:2022sfx} will be discussed
  elsewhere.}\BibitemShut {Stop}%
\bibitem [{Note2()}]{Note2}%
  \BibitemOpen
  \bibinfo {note} {We set the radius $L$ of an asymptotic $AdS_4$ geometry to
  unity.}\BibitemShut {Stop}%
\bibitem [{\citenamefont {Buchel}\ and\ \citenamefont
  {Pagnutti}(2010)}]{Buchel:2009ge}%
  \BibitemOpen
  \bibfield  {author} {\bibinfo {author} {\bibfnamefont {A.}~\bibnamefont
  {Buchel}}\ and\ \bibinfo {author} {\bibfnamefont {C.}~\bibnamefont
  {Pagnutti}},\ }\href {\doibase 10.1016/j.nuclphysb.2009.08.017} {\bibfield
  {journal} {\bibinfo  {journal} {Nucl. Phys. B}\ }\textbf {\bibinfo {volume}
  {824}},\ \bibinfo {pages} {85} (\bibinfo {year} {2010})},\ \Eprint
  {http://arxiv.org/abs/0904.1716} {arXiv:0904.1716 [hep-th]} \BibitemShut
  {NoStop}%
\bibitem [{\citenamefont {Bosch}\ \emph {et~al.}(2017)\citenamefont {Bosch},
  \citenamefont {Buchel},\ and\ \citenamefont {Lehner}}]{Bosch:2017ccw}%
  \BibitemOpen
  \bibfield  {author} {\bibinfo {author} {\bibfnamefont {P.}~\bibnamefont
  {Bosch}}, \bibinfo {author} {\bibfnamefont {A.}~\bibnamefont {Buchel}}, \
  and\ \bibinfo {author} {\bibfnamefont {L.}~\bibnamefont {Lehner}},\ }\href
  {\doibase 10.1007/JHEP07(2017)135} {\bibfield  {journal} {\bibinfo  {journal}
  {JHEP}\ }\textbf {\bibinfo {volume} {07}},\ \bibinfo {pages} {135} (\bibinfo
  {year} {2017})},\ \Eprint {http://arxiv.org/abs/1704.05454} {arXiv:1704.05454
  [hep-th]} \BibitemShut {NoStop}%
\bibitem [{\citenamefont {Buchel}(2022{\natexlab{b}})}]{Buchel:2022hjz}%
  \BibitemOpen
  \bibfield  {author} {\bibinfo {author} {\bibfnamefont {A.}~\bibnamefont
  {Buchel}},\ }\href {\doibase 10.1007/JHEP09(2022)227} {\bibfield  {journal}
  {\bibinfo  {journal} {JHEP}\ }\textbf {\bibinfo {volume} {09}},\ \bibinfo
  {pages} {227} (\bibinfo {year} {2022}{\natexlab{b}})},\ \Eprint
  {http://arxiv.org/abs/2207.09887} {arXiv:2207.09887 [hep-th]} \BibitemShut
  {NoStop}%
\bibitem [{\citenamefont {Chesler}\ and\ \citenamefont
  {Yaffe}(2014)}]{Chesler:2013lia}%
  \BibitemOpen
  \bibfield  {author} {\bibinfo {author} {\bibfnamefont {P.~M.}\ \bibnamefont
  {Chesler}}\ and\ \bibinfo {author} {\bibfnamefont {L.~G.}\ \bibnamefont
  {Yaffe}},\ }\href {\doibase 10.1007/JHEP07(2014)086} {\bibfield  {journal}
  {\bibinfo  {journal} {JHEP}\ }\textbf {\bibinfo {volume} {07}},\ \bibinfo
  {pages} {086} (\bibinfo {year} {2014})},\ \Eprint
  {http://arxiv.org/abs/1309.1439} {arXiv:1309.1439 [hep-th]} \BibitemShut
  {NoStop}%
\bibitem [{\citenamefont {Booth}(2005)}]{Booth:2005qc}%
  \BibitemOpen
  \bibfield  {author} {\bibinfo {author} {\bibfnamefont {I.}~\bibnamefont
  {Booth}},\ }\href {\doibase 10.1139/p05-063} {\bibfield  {journal} {\bibinfo
  {journal} {Can. J. Phys.}\ }\textbf {\bibinfo {volume} {83}},\ \bibinfo
  {pages} {1073} (\bibinfo {year} {2005})},\ \Eprint
  {http://arxiv.org/abs/gr-qc/0508107} {arXiv:gr-qc/0508107 [gr-qc]}
  \BibitemShut {NoStop}%
\bibitem [{\citenamefont {Figueras}\ \emph {et~al.}(2009)\citenamefont
  {Figueras}, \citenamefont {Hubeny}, \citenamefont {Rangamani},\ and\
  \citenamefont {Ross}}]{Figueras:2009iu}%
  \BibitemOpen
  \bibfield  {author} {\bibinfo {author} {\bibfnamefont {P.}~\bibnamefont
  {Figueras}}, \bibinfo {author} {\bibfnamefont {V.~E.}\ \bibnamefont
  {Hubeny}}, \bibinfo {author} {\bibfnamefont {M.}~\bibnamefont {Rangamani}}, \
  and\ \bibinfo {author} {\bibfnamefont {S.~F.}\ \bibnamefont {Ross}},\ }\href
  {\doibase 10.1088/1126-6708/2009/04/137} {\bibfield  {journal} {\bibinfo
  {journal} {JHEP}\ }\textbf {\bibinfo {volume} {04}},\ \bibinfo {pages} {137}
  (\bibinfo {year} {2009})},\ \Eprint {http://arxiv.org/abs/0902.4696}
  {arXiv:0902.4696 [hep-th]} \BibitemShut {NoStop}%
\bibitem [{\citenamefont {Buchel}(2017{\natexlab{b}})}]{Buchel:2017map}%
  \BibitemOpen
  \bibfield  {author} {\bibinfo {author} {\bibfnamefont {A.}~\bibnamefont
  {Buchel}},\ }\href {\doibase 10.1007/JHEP08(2017)134} {\bibfield  {journal}
  {\bibinfo  {journal} {JHEP}\ }\textbf {\bibinfo {volume} {08}},\ \bibinfo
  {pages} {134} (\bibinfo {year} {2017}{\natexlab{b}})},\ \Eprint
  {http://arxiv.org/abs/1705.08560} {arXiv:1705.08560 [hep-th]} \BibitemShut
  {NoStop}%
\bibitem [{Note3()}]{Note3}%
  \BibitemOpen
  \bibinfo {note} {In practice we initialize the system in an arbitrary
  spatially homogeneous and isotropic state at $Ht_{init}\ll -\protect \frac
  {5}{\gamma }H$ and let it evolve to an appropriate de Sitter DFP attractor,
  uniquely specified by the ratio $\protect \frac {\Lambda }{H}$ \cite
  {Buchel:2017lhu}.}\BibitemShut {Stop}%
\bibitem [{Note4()}]{Note4}%
  \BibitemOpen
  \bibinfo {note} {Additional details regarding the simulations can be found in
  the supplemental material.}\BibitemShut {Stop}%
\bibitem [{\citenamefont {Buchel}\ \emph {et~al.}(2015)\citenamefont {Buchel},
  \citenamefont {Heller},\ and\ \citenamefont {Myers}}]{Buchel:2015saa}%
  \BibitemOpen
  \bibfield  {author} {\bibinfo {author} {\bibfnamefont {A.}~\bibnamefont
  {Buchel}}, \bibinfo {author} {\bibfnamefont {M.~P.}\ \bibnamefont {Heller}},
  \ and\ \bibinfo {author} {\bibfnamefont {R.~C.}\ \bibnamefont {Myers}},\
  }\href {\doibase 10.1103/PhysRevLett.114.251601} {\bibfield  {journal}
  {\bibinfo  {journal} {Phys. Rev. Lett.}\ }\textbf {\bibinfo {volume} {114}},\
  \bibinfo {pages} {251601} (\bibinfo {year} {2015})},\ \Eprint
  {http://arxiv.org/abs/1503.07114} {arXiv:1503.07114 [hep-th]} \BibitemShut
  {NoStop}%
\bibitem [{\citenamefont {Buchel}\ and\ \citenamefont
  {Day}(2015)}]{Buchel:2015ofa}%
  \BibitemOpen
  \bibfield  {author} {\bibinfo {author} {\bibfnamefont {A.}~\bibnamefont
  {Buchel}}\ and\ \bibinfo {author} {\bibfnamefont {A.}~\bibnamefont {Day}},\
  }\href {\doibase 10.1103/PhysRevD.92.026009} {\bibfield  {journal} {\bibinfo
  {journal} {Phys. Rev. D}\ }\textbf {\bibinfo {volume} {92}},\ \bibinfo
  {pages} {026009} (\bibinfo {year} {2015})},\ \Eprint
  {http://arxiv.org/abs/1505.05012} {arXiv:1505.05012 [hep-th]} \BibitemShut
  {NoStop}%
\bibitem [{\citenamefont {Attems}\ \emph {et~al.}(2016)\citenamefont {Attems},
  \citenamefont {Casalderrey-Solana}, \citenamefont {Mateos}, \citenamefont
  {Papadimitriou}, \citenamefont {Santos-Oliv\'an}, \citenamefont {Sopuerta},
  \citenamefont {Triana},\ and\ \citenamefont {Zilh\~ao}}]{Attems:2016ugt}%
  \BibitemOpen
  \bibfield  {author} {\bibinfo {author} {\bibfnamefont {M.}~\bibnamefont
  {Attems}}, \bibinfo {author} {\bibfnamefont {J.}~\bibnamefont
  {Casalderrey-Solana}}, \bibinfo {author} {\bibfnamefont {D.}~\bibnamefont
  {Mateos}}, \bibinfo {author} {\bibfnamefont {I.}~\bibnamefont
  {Papadimitriou}}, \bibinfo {author} {\bibfnamefont {D.}~\bibnamefont
  {Santos-Oliv\'an}}, \bibinfo {author} {\bibfnamefont {C.~F.}\ \bibnamefont
  {Sopuerta}}, \bibinfo {author} {\bibfnamefont {M.}~\bibnamefont {Triana}}, \
  and\ \bibinfo {author} {\bibfnamefont {M.}~\bibnamefont {Zilh\~ao}},\ }\href
  {\doibase 10.1007/JHEP10(2016)155} {\bibfield  {journal} {\bibinfo  {journal}
  {JHEP}\ }\textbf {\bibinfo {volume} {10}},\ \bibinfo {pages} {155} (\bibinfo
  {year} {2016})},\ \Eprint {http://arxiv.org/abs/1603.01254} {arXiv:1603.01254
  [hep-th]} \BibitemShut {NoStop}%
\bibitem [{\citenamefont {Janik}\ \emph {et~al.}(2016)\citenamefont {Janik},
  \citenamefont {Jankowski},\ and\ \citenamefont
  {Soltanpanahi}}]{Janik:2016btb}%
  \BibitemOpen
  \bibfield  {author} {\bibinfo {author} {\bibfnamefont {R.~A.}\ \bibnamefont
  {Janik}}, \bibinfo {author} {\bibfnamefont {J.}~\bibnamefont {Jankowski}}, \
  and\ \bibinfo {author} {\bibfnamefont {H.}~\bibnamefont {Soltanpanahi}},\
  }\href {\doibase 10.1007/JHEP06(2016)047} {\bibfield  {journal} {\bibinfo
  {journal} {JHEP}\ }\textbf {\bibinfo {volume} {06}},\ \bibinfo {pages} {047}
  (\bibinfo {year} {2016})},\ \Eprint {http://arxiv.org/abs/1603.05950}
  {arXiv:1603.05950 [hep-th]} \BibitemShut {NoStop}%
\bibitem [{Note5()}]{Note5}%
  \BibitemOpen
  \bibinfo {note} {It can be analytically computed using the near-conformal
  description of the de Sitter DFPs of the model \cite {Buchel:2017lhu} and its
  equilibrium thermodynamics.}\BibitemShut {Stop}%
\bibitem [{Note6()}]{Note6}%
  \BibitemOpen
  \bibinfo {note} {See \cite {Akrami:2017cir} for a recent review and
  additional references.}\BibitemShut {Stop}%
\bibitem [{\citenamefont {Peebles}\ and\ \citenamefont
  {Vilenkin}(1999)}]{Peebles:1998qn}%
  \BibitemOpen
  \bibfield  {author} {\bibinfo {author} {\bibfnamefont {P.~J.~E.}\
  \bibnamefont {Peebles}}\ and\ \bibinfo {author} {\bibfnamefont
  {A.}~\bibnamefont {Vilenkin}},\ }\href {\doibase 10.1103/PhysRevD.59.063505}
  {\bibfield  {journal} {\bibinfo  {journal} {Phys. Rev. D}\ }\textbf {\bibinfo
  {volume} {59}},\ \bibinfo {pages} {063505} (\bibinfo {year} {1999})},\
  \Eprint {http://arxiv.org/abs/astro-ph/9810509} {arXiv:astro-ph/9810509}
  \BibitemShut {NoStop}%
\bibitem [{\citenamefont {Felder}\ \emph {et~al.}(1999)\citenamefont {Felder},
  \citenamefont {Kofman},\ and\ \citenamefont {Linde}}]{Felder:1999pv}%
  \BibitemOpen
  \bibfield  {author} {\bibinfo {author} {\bibfnamefont {G.~N.}\ \bibnamefont
  {Felder}}, \bibinfo {author} {\bibfnamefont {L.}~\bibnamefont {Kofman}}, \
  and\ \bibinfo {author} {\bibfnamefont {A.~D.}\ \bibnamefont {Linde}},\ }\href
  {\doibase 10.1103/PhysRevD.60.103505} {\bibfield  {journal} {\bibinfo
  {journal} {Phys. Rev. D}\ }\textbf {\bibinfo {volume} {60}},\ \bibinfo
  {pages} {103505} (\bibinfo {year} {1999})},\ \Eprint
  {http://arxiv.org/abs/hep-ph/9903350} {arXiv:hep-ph/9903350} \BibitemShut
  {NoStop}%
\bibitem [{Note7()}]{Note7}%
  \BibitemOpen
  \bibinfo {note} {This is a different universality from the abrupt holographic
  quenches of the relevant couplings of the boundary QFT studied in \cite
  {Buchel:2013gba}.}\BibitemShut {Stop}%
\bibitem [{Note8()}]{Note8}%
  \BibitemOpen
  \bibinfo {note} {We do not provide the plot for the comoving entropy density
  for the evolution with $\protect \frac {\gamma }{H}=10^{-1}$. Given the vast
  expansion of the Universe in this case, $\protect \qopname \relax
  o{ln}\protect \frac {a_e}{a_s}=50$, it would completely overwhelm the other
  results.}\BibitemShut {Stop}%
\bibitem [{\citenamefont {Buchel}(2019)}]{Buchel:2019qcq}%
  \BibitemOpen
  \bibfield  {author} {\bibinfo {author} {\bibfnamefont {A.}~\bibnamefont
  {Buchel}},\ }\href {\doibase 10.1016/j.nuclphysb.2019.114769} {\bibfield
  {journal} {\bibinfo  {journal} {Nucl. Phys. B}\ }\textbf {\bibinfo {volume}
  {948}},\ \bibinfo {pages} {114769} (\bibinfo {year} {2019})},\ \Eprint
  {http://arxiv.org/abs/1904.09968} {arXiv:1904.09968 [hep-th]} \BibitemShut
  {NoStop}%
\bibitem [{\citenamefont {Buchel}(2022{\natexlab{c}})}]{Buchel:2022sfx}%
  \BibitemOpen
  \bibfield  {author} {\bibinfo {author} {\bibfnamefont {A.}~\bibnamefont
  {Buchel}},\ }\href@noop {} {\  (\bibinfo {year} {2022}{\natexlab{c}})},\
  \Eprint {http://arxiv.org/abs/2210.17380} {arXiv:2210.17380 [hep-th]}
  \BibitemShut {NoStop}%
\bibitem [{\citenamefont {Akrami}\ \emph {et~al.}(2018)\citenamefont {Akrami},
  \citenamefont {Kallosh}, \citenamefont {Linde},\ and\ \citenamefont
  {Vardanyan}}]{Akrami:2017cir}%
  \BibitemOpen
  \bibfield  {author} {\bibinfo {author} {\bibfnamefont {Y.}~\bibnamefont
  {Akrami}}, \bibinfo {author} {\bibfnamefont {R.}~\bibnamefont {Kallosh}},
  \bibinfo {author} {\bibfnamefont {A.}~\bibnamefont {Linde}}, \ and\ \bibinfo
  {author} {\bibfnamefont {V.}~\bibnamefont {Vardanyan}},\ }\href {\doibase
  10.1088/1475-7516/2018/06/041} {\bibfield  {journal} {\bibinfo  {journal}
  {JCAP}\ }\textbf {\bibinfo {volume} {06}},\ \bibinfo {pages} {041} (\bibinfo
  {year} {2018})},\ \Eprint {http://arxiv.org/abs/1712.09693} {arXiv:1712.09693
  [hep-th]} \BibitemShut {NoStop}%
\bibitem [{\citenamefont {Buchel}\ \emph {et~al.}(2013)\citenamefont {Buchel},
  \citenamefont {Myers},\ and\ \citenamefont {van Niekerk}}]{Buchel:2013gba}%
  \BibitemOpen
  \bibfield  {author} {\bibinfo {author} {\bibfnamefont {A.}~\bibnamefont
  {Buchel}}, \bibinfo {author} {\bibfnamefont {R.~C.}\ \bibnamefont {Myers}}, \
  and\ \bibinfo {author} {\bibfnamefont {A.}~\bibnamefont {van Niekerk}},\
  }\href {\doibase 10.1103/PhysRevLett.111.201602} {\bibfield  {journal}
  {\bibinfo  {journal} {Phys. Rev. Lett.}\ }\textbf {\bibinfo {volume} {111}},\
  \bibinfo {pages} {201602} (\bibinfo {year} {2013})},\ \Eprint
  {http://arxiv.org/abs/1307.4740} {arXiv:1307.4740 [hep-th]} \BibitemShut
  {NoStop}%
\end{thebibliography}%

\end{document}